\documentclass[11pt,onecolumn]{IEEEtran}
\usepackage[utf8]{inputenc} 
\usepackage[T1]{fontenc}
\usepackage{url}
\usepackage{ifthen}
\usepackage{cite}
\usepackage[cmex10]{amsmath} 
\usepackage{graphicx}
\usepackage{amsthm,amssymb,amsfonts}
\usepackage{xcolor}
\usepackage{enumitem}
\usepackage{multirow}
\usepackage{caption}

\newcommand{\bit}{\begin{itemize}}
	\newcommand{\eit}{\end{itemize}}
\newcommand{\bcor}{\begin{cor}}
	\newcommand{\ecor}{\end{cor}}
\newcommand{\beq}{\begin{equation}}
\newcommand{\eeq}{\end{equation}}
\newcommand{\beqn}{\begin{equation*}}
\newcommand{\eeqn}{\end{equation*}}
\newcommand{\bea}{\begin{eqnarray}}
\newcommand{\eea}{\end{eqnarray}}
\newcommand{\bean}{\begin{eqnarray*}}
	\newcommand{\eean}{\end{eqnarray*}}
\newcommand{\ben}{\begin{enumerate}}
	\newcommand{\een}{\end{enumerate}}
\newcommand{\bdefn}{\begin{defn}}
	\newcommand{\edefn}{\end{defn}}
\newcommand{\bnote}{\begin{note}}
	\newcommand{\enote}{\end{note}}
\newcommand{\bprop}{\begin{prop}}
	\newcommand{\eprop}{\end{prop}}
\newcommand{\blem}{\begin{lem}}
	\newcommand{\elem}{\end{lem}}
\newcommand{\bthm}{\begin{thm}}
	\newcommand{\ethm}{\end{thm}}
\newcommand{\bconj}{\begin{conj}}
	\newcommand{\econj}{\end{conj}}
\newcommand{\bconstr}{\begin{constr}}
	\newcommand{\econstr}{\end{constr}}
\newcommand{\bpf}{\begin{proof}}
	\newcommand{\epf}{\end{proof}}
\newcommand{\bc}{\begin{center}}
	\newcommand{\ec}{\end{center}}
\newcommand{\bprf}{{\em Proof: }}
\newcommand{\eprf}{\hfill $\Box$}

\newtheorem{thm}{Theorem}[section]

\newtheorem{lem}[thm]{Lemma}
\newtheorem{cor}[thm]{Corollary}
\newtheorem{defn}{Definition}[section]
\newtheorem{note}{Remark}[section]

\newcommand{\params}{\mbox{$(a,b,\tau)$}}

\newtheorem{constr}[thm]{Construction}

\IEEEoverridecommandlockouts 
\begin{document}
	
\title{Staggered Diagonal Embedding Based \\ Linear Field Size Streaming Codes} 
 \author{%
	\IEEEauthorblockN{Vinayak Ramkumar\IEEEauthorrefmark{1},
		Myna Vajha\IEEEauthorrefmark{1},
		M. Nikhil Krishnan\IEEEauthorrefmark{2}, P. Vijay Kumar\IEEEauthorrefmark{1}\\}
	\IEEEauthorblockA{\IEEEauthorrefmark{1}%
		Department of Electrical Communication Engineering, IISc Bangalore\\}
	\IEEEauthorblockA{\IEEEauthorrefmark{2}%
		Department of Electrical and Computer Engineering, University of Toronto\\}
	\IEEEauthorblockA{
		\{vinram93, mynaramana, nikhilkrishnan.m, pvk1729\}@gmail.com}
	\thanks{All the authors contributed equally to this work. P. Vijay Kumar is also a  Visiting  Professor  at  the  University  of  Southern  California. This  research  is  supported  in  part  by  the J C Bose National Fellowship JCB/2017/000017 and  in  part  by the  NetApp University Research Fund SVCF-0002.  Myna Vajha would like  to  acknowledge  the  support  of  Visvesvaraya  PhD  Scheme.  Myna Vajha and M. Nikhil Krishnan would like to acknowledge the support of Qualcomm Innovation Fellowship, India.}
}

\maketitle

\begin{abstract}
	
An $(a,b,\tau)$ streaming code is a packet-level erasure code that can recover under a strict delay constraint of $\tau$ time units, from either a burst of $b$ erasures or else of $a$ random erasures, occurring within a sliding window of time duration $w$. While rate-optimal constructions of such streaming codes are available for all parameters $\{a,b,\tau,w\}$ in the literature, they require in most instances, a quadratic, $O(\tau^2)$ field size. In this work, we make further progress towards field size reduction and present rate-optimal $O(\tau)$ field size streaming codes for two regimes: (i) $gcd(b,\tau+1-a)\ge a$ (ii) $\tau+1 \ge a+b$ and  $b \mod \ a \in \{0,a-1\}$.     
 
\end{abstract}



\section{Introduction}
Ultra-Reliable, Low-Latency Communication (URLLC) is a principal focus area of 5G and is key to enabling many next-generation applications such as interactive streaming, industrial automation, multi-player gaming and disaster recovery. ARQ-based schemes, while ensuring reliability, are not suitable for low-latency communication due to their large round-trip delays.  The naive solution of replication to ensure reliability leads to wastage of resources. Thus, the development of  FEC schemes that can operate under a strict decoding-delay constraint is necessary for the setting up of a reliable, low-latency communication system. The streaming codes under discussion here, were developed with this aim in mind.      

In \cite{MartSunTIT04} and \cite{MartTrotISIT07}, authors introduced the setting of streaming codes, which is as follows. There is an infinite stream of message packets $\{\underline{u}(t)\}_{t=0}^{^\infty}$, $\underline{u}(t) \in \mathbb{F}_q^k$, which needs to be reliably transmitted from a transmitter to a receiver, where the channel can introduce packet losses. In order to tackle packet losses, coded packets which contain both message and parity parts are transmitted across the channel. We use the terminology message packet to denote the message part of the coded packet and similarly parity packet refers to the parity part of the coded packet. Let $\underline{x}(t) \in \mathbb{F}_q^n$ denote the coded packet transmitted at time $t$. Then $\underline{x}(t)^T \triangleq \left[\underline{u}(t)^T \ \underline{p}(t)^T\right]$, where $\underline{u}(t) \in \mathbb{F}_q^k$ is the message packet at time $t$  and $\underline{p}(t) \in \mathbb{F}_q^{n-k}$ is the parity packet at time $t$. The parity packet $\underline{p}(t)$  at time $t$ is a function only of $\{\underline{u}(\ell)~|~\ell \le t\}$, due to the causal nature of the encoder. The initial channel model considered for streaming codes in \cite{MartSunTIT04} and \cite{MartTrotISIT07} is such that in every sliding window of time duration $\tau+1$, there can be a burst erasure of length at most $b$. Streaming code constructions are presented in \cite{MartSunTIT04} and \cite{MartTrotISIT07}, which permit recovery of each message packet with a delay of at most $\tau$, in spite of the burst losses, i.e., $\underline{u}(t)$ is recovered by time $t+\tau$, for all $t$. In a subsequent work, Badr et al. \cite{BadrPatilKhistiTIT17} introduced the delay-constrained sliding-window (DC-SW) channel model, which is a tractable deterministic approximation of the popularly used Gilbert-Elliott (GE) channel model. Under the DC-SW channel model, within any sliding window of time duration $w$, there can be either at most $a$ random erasures or else, a burst erasure of length $\leq b$. The paper \cite{BadrPatilKhistiTIT17} presented streaming code constructions which can recover every packet $\underline{u}(t)$ by time $t+\tau$ in presence of the DC-SW channel. It is to be noted that the channel parameters naturally satisfy: $a \le b \le \tau$.   Without loss of generality, one can set $w=\tau+1$ (see \cite{BadrPatilKhistiTIT17} or \cite{NikDeepPVK}).  Hence the DC-SW channel is parameterized by $\{a,b,\tau\}$. In the remainder of the paper, we use \params\ streaming code to refer to codes which can recover from all the permissible erasure patterns of $\{a,b,\tau\}$ DC-SW channel, under strict decoding delay constraint $\tau$.
 
\begin{figure}[htbp]
	\centering
	\includegraphics[scale=0.35]{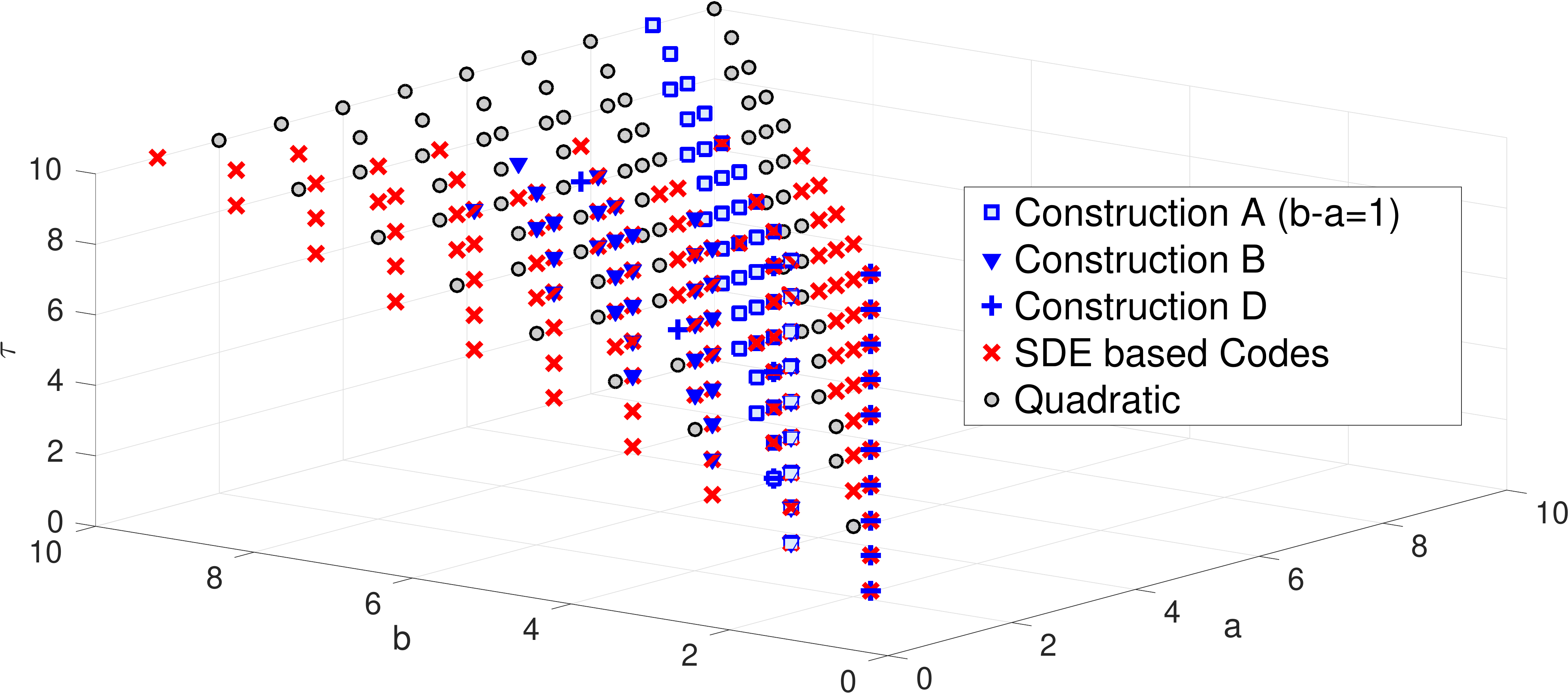}
	\caption{The figure depicts for all parameters $\{a,b,\tau\}$, where $a\le b \le \tau \le 10$, the smallest field size streaming code known. The construction A ($b-a=1$), B and D refer to the linear field size codes from \cite{NikDeepPVK} and SDE-based codes are the linear field size codes presented in this paper. For the rest of the valid parameters, best known codes require quadratic field size and can be found in \cite{NikDeepPVK}, \cite{KhistiExplicitCode}. }
		\label{fig:3d_plot}
\end{figure}

In \cite{BadrPatilKhistiTIT17}, an upper bound on the rate $R$ of an \params\ streaming code is provided. In \cite{FongKhistiTIT19}, \cite{NikPVK} it was shown that this upper bound is indeed achievable for all parameters. The optimal rate of an \params\ streaming code thus obtained is given by, \bean
R_\text{opt} \triangleq  \frac{\tau+1-a}{\tau+1-a+b}.
\eean
The papers \cite{FongKhistiTIT19,NikPVK} presented first families of rate-optimal streaming code constructions and required a field size exponential in $\tau$. In  \cite{NikDeepPVK}, an $O(\tau^2)$ field size non-explicit rate-optimal streaming code construction is presented for all possible $\{a,b,\tau\}$. The paper \cite{NikDeepPVK} also provided $4$ additional constructions with $O(\tau)$ field size for restricted parameter sets (see Table~\ref{tab:para}). An explicit quadratic field size streaming code construction for all parameters is presented in \cite{KhistiExplicitCode}. The rate-optimal streaming code constructions appearing in \cite{MartTrotISIT07,FongKhistiTIT19,NikPVK,NikDeepPVK,KhistiExplicitCode} all employ a certain  diagonal embedding (DE)  technique introduced in   \cite{MartTrotISIT07}. The DE technique enables one to construct streaming codes by diagonally embedding the code symbols of a scalar block code in the packet stream. In a recent work \cite{simple}, the authors introduced the technique of staggered diagonal embedding (SDE), which generalizes DE. 
  Under the SDE approach, $n$ code symbols of the scalar block code are dispersed across a span of $N \ge n$ successive packets. A linear field size rate-optimal streaming code construction for all $\{a,b,\tau\}$ such that $gcd(b,\tau+1-a) = b$ is presented in \cite{simple}. The study of SDE in \cite{simple} is restricted to the case $N \le \tau+1$. In the present paper, we explore the SDE technique beyond $N \le \tau+1$ and present new streaming code constructions, which require $O(\tau)$ field size and smaller packet length $n$, compared to existing constructions. As shown in Fig. \ref{fig:3d_plot}, constructions in the present paper provide a significant range of new parameters $\{a,b,\tau\}$ over which linear field size is feasible.
   
\subsubsection*{Our Contributions}
\bit
\item We provide necessary and sufficient conditions for SDE of a scalar code to result in an \params\ streaming code. 
\item We develop a new family of scalar codes which result in linear field size, rate-optimal \params\ streaming codes for two new regimes.
\bit
 \item We use SDE to generate \params\ streaming codes for all $\{a,b,\tau\}$ with $gcd(b,\tau+1-a) \ge a$. 
\item We show using DE (a special case of SDE where $n=N$) that the scalar code construction results in \params\ streaming codes whenever $\tau+1 \ge a+b$ and  $b \mod \ a = 0$ and a modified version of this scalar code works whenever $\tau+1 \ge a+b$ and  $b \mod \ a = a-1$.
\eit     
\eit
\subsubsection*{Organization of the Paper}
 The SDE framework is described in full generality in  Section~\ref{sec:sde}. In Section~\ref{sec:building}, we  provide the construction of scalar block code to be used in conjunction with SDE or the special case of DE. We provide construction of linear field size streaming codes based on SDE technique in Section~\ref{sec:constr}. In Section \ref{sec:de_constr}, we show how the construction can be modified to come up with DE-based linear field size streaming codes. 

\subsubsection*{Notation}
We use the notation $[a:b]$ to denote $\{a,a+1, \dots, b-1, b\}$. For any finite set $E \subseteq \mathbb{Z}$, we use $|E|$ to denote number of elements in $E$ and $\max(E)$ to denote the largest element in $E$. Let $M \in \mathbb{F}_q^{k \times n}$, then by $M(i_1:i_2, j_1:j_2)$, we mean the sub-matrix of $M$  comprising of rows whose indices lie in $[i_1:i_2]$ and columns whose indices lie in $[j_1:j_2]$. 
We use the notation $I_k$ for the $k \times k$ identity matrix. 
For $V \subseteq \mathbb{F}_q^n$, span$\langle V\rangle$ denotes the linear span of $V$.

\begin{table}[h!]

	\begin{center}
		\bean
		\begin{array}{|c|c|} \hline
			\text{Linear Field Size Streaming Code} & \text{Parameter Range}  \\ \hline \hline   
			\text{Construction A \cite{NikDeepPVK}} & b-a=1  \\ \hline 
			\text{Construction B \cite{NikDeepPVK}} & (\tau+a+1) \ge 2b \ge 4a \\ \hline 
			\text{Construction C \cite{NikDeepPVK}} & a | b| (\tau+1-a) \\ \hline 
			\text{Construction D \cite{NikDeepPVK}} &  b=2a-1 \text{ and } b|(\tau+2-a)\\ \hline 
			\text{Simple Streaming Code \cite{simple}} &  b|(\tau+1-a)\\ \hline
			\text{SDE-based code (present paper)} &  gcd(b,\tau+1-a)\ge a\\ \hline  
			\text{DE-based code (present paper)} &  \tau+1-a \ge b \text{ and } \\ &  b \mod a \in \{0,a-1\}\\ \hline 
		\end{array}
		\eean
		\caption{Parameters for which linear field size streaming codes are known.}	
		\label{tab:para}
	\end{center}
\end{table}

\section{Staggered Diagonal Embedding} \label{sec:sde}
\begin{figure}[htbp]
	\centering
	\includegraphics[width=0.7\textwidth]{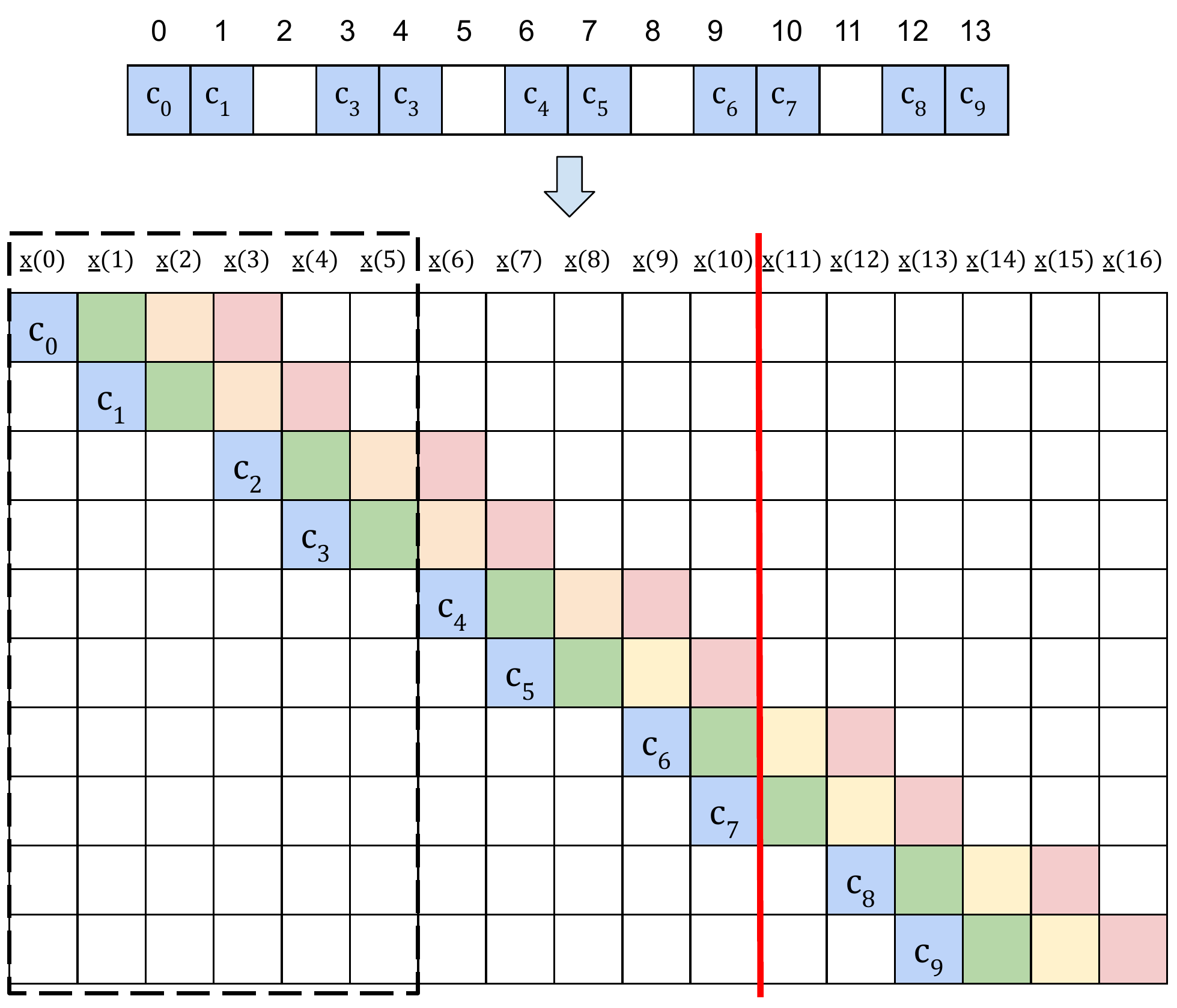}
	\caption{SDE of an $[n=10,k=6]$ base code with placement set $S=\{0,1,3,4,6,7,9,10,12,13\}$. Here $(c_0\ c_1\ \ldots\ c_{9})$ is a codeword in the base code $\mathbb{C}$. It will be shown in Section~\ref{sec:constr} that this results in an $(a=2,b=6,\tau=10)$ streaming code. The black dashed window indicates a burst of size $6$ starting at index $0$ and the red line indicates the decoding delay i.e, recovery of $0$th packet can access packets until index $10$. }
	\label{fig:sde}
\end{figure}

In this section, we explain the staggered diagonal embedding technique introduced in \cite{simple}, for constructing packet-level codes from scalar codes. Let $\mathbb{C}$ be an $[n,k]$ linear code over $\mathbb{F}_q$, with first $k$ symbols forming an information set. Let $N \ge n$ be an integer and let $S \subseteq [0:N-1]$ be such that $|S|=n$. Let $S\triangleq\{s_0, s_1, \dots , s_{n-1}\}$, where $0=s_0 < s_1 < \dots < s_{n-1}=N-1$. We refer to $\mathbb{C}$ as the base code, $S$ as the placement set and $N$ as the dispersion span. The packet-level code resulting from SDE of scalar code $\mathbb{C}$ with the placement set $S$ will be referred to as SDE$(\mathbb{C}, S)$. For $i\in[0:n-1]$, let $x_i(t)$ denote the $i$th component of the coded packet $\underline{x}(t)$ of the packet-level code SDE$(\mathbb{C}, S)$ (see Fig.~\ref{fig:sde} for an example). 
Then we have the following relation between component symbols: 
\bea
\label{eq:sde}\Big(x_0(t+s_0), x_1(t+s_1),\cdots, x_{n-1}(t+s_{n-1})\Big) \in \mathbb{C}, ~\forall t.
\eea 
It is easy to see that the resultant packet-level code has rate $\frac{k}{n}$, which is same as that of $\mathbb{C}$. When $N=n$, we have $S=[0:n-1]$ and SDE reduces to DE.

For the packet-level code SDE$(\mathbb{C},S)$ to be an $(a,b,\tau)$ streaming code, there are some conditions that it needs to satisfy. The Theorem $1$ in \cite{simple} states such conditions for the case $N\le \tau +1$, whereas here in Theorem~\ref{thm:sde}, we provide necessary and sufficient conditions for the general case which includes $N>\tau+1$. 

For a streaming code, a lost packet $\underline{x}(t)$ must be recovered from admissible erasures by accessing all the available packets till time  $t+\tau$. We need to translate this requirement in terms of the scalar code $\mathbb{C}$. Towards this, we first introduce the function $f_S:[0:N-1] \rightarrow [0:n-1]$
for a given placement set $S$, which is defined as:
\beqn
f_S(j)\triangleq \max\{i:s_i\leq j\}.
\eeqn
We now define $r_i \triangleq f_S(\min\{s_i+\tau,N-1\})$ for every index $i \in [0: n-1]$. During the recovery of $i$th code symbol, one can access only till $r_i$th code symbol. We will use the notation $f_S(J)$ to indicate the set $\{f_S(j)~|~j \in J\}$.

As an example, in Fig.~\ref{fig:sde}, we have $n=10,N=14, \tau=10$ and $S=\{0,1,3,4,6,7,9,10,12,13\}$. In order to recover packet $\underline{x}(t)$ from a burst of size $6$ starting at time $t$, one can access packets only till time $t+10$. In terms of $\mathbb{C}$, as we have $r_0 = f_S(10)=7$, for recovering $c_0$, symbols only till $c_7$ are accessible. 

\begin{thm}\label{thm:sde}
	Let  $\mathbb{C}$ be an $[n,k]$ base code over $\mathbb{F}_q$ and let $S=\{s_0, s_1, \dots , s_{n-1}\} \subseteq [0:N-1]$ be a placement set, where $s_0=0<s_1\dots<s_{n-1}=N-1$. The packet-level code SDE$(\mathbb{C}, S)$ is an $(a, b, \tau)$ streaming code iff $\mathbb{C}, S$ satisfy the following conditions: 
	\begin{enumerate}
\item Random erasure recovery : for all $i \in [0:n-1]$ and  every $A \subseteq [i:r_i]$ such that $i \in A$, $|A| \le a$,  $c_i $ is a linear combination of $\{c_j ~|~ j<i\} \cup \{c_j ~|~ j \in [i:r_i] \setminus A\}$ over $\mathbb{F}_q$, for all  $\underline{c}=(c_0,c_1,\dots,c_{n-1}) \in \mathbb{C}$.
	\item[]  
\item Burst erasure recovery: for all $i \in [0:n-1]$ and every $B \subseteq [s_i:\min\{s_i+b-1,N-1\}]$ such that $s_i \in B$,  $c_i $ is a linear combination of $\{c_j ~|~ j<i\} \cup \{c_j ~|~ j \in [i:r_i] \setminus f_S(B)\}$ over $\mathbb{F}_q$, for all  $\underline{c}=(c_0,c_1,\dots,c_{n-1}) \in \mathbb{C}$.
	\end{enumerate}
\end{thm}
\bprf 
By definition, the packet-level code SDE$(\mathbb{C},S)$ is an $(a,b,\tau)$ streaming code iff for every $t$ and every $E \subseteq [t: t+\tau]$ such that $t \in E$ and either $|E| \le a$ or $\max E - t \le b-1$, the packet $\underline{x}(t)$ is recoverable from $\{x(t') \mid t' < t \text{~or~} t' \in [t: t+\tau] \setminus E\}$. Let $x_i(t)$ be the $i$-th component of the packet $\underline{x}(t)$. Since the packets are coded using SDE$(\mathbb{C}, S)$, by \eqref{eq:sde}, the following condition is satisfied for every $i \in [0: n-1]$:
\bean
	(x_0(t-s_i+s_0), \cdots, x_i(t), \cdots, x_{n-1}(t-s_i+s_{n-1}))\triangleq \underline{c} \in \mathbb{C}.
\eean
For any erasure set $E \subseteq [t: t+\tau] $ such that $t \in E$ and $|E| \le a$, the codeword $\underline{c}$ observes erasures across coordinates indexed by $A \triangleq f_S(\{e+s_i-t~|~e \in E,~e+s_i-t \le N-1\}) \subseteq [i:r_i]$. For any such erasure $E$, the symbols $x_i(t)\triangleq c_i$ for all $i \in [0:n-1]$, can be recovered, iff condition 1 holds. Now, for any erasure set $E \subseteq [t: t+\tau]$ such that $t \in E$, $\max(E)-t \le b-1$, let $B \triangleq \{e+s_i-t~|~e \in E,~e+s_i-t \le N-1\}$. The erasures that the codeword $\underline{c}$ observes are given by $f_S(B)$ and the recovery of symbols $x_i(t)$ for all $i \in [0:n-1]$ is ensured iff condition $2$ holds.
Therefore conditions 1 and 2 are necessary and sufficient conditions for SDE$(\mathbb{C}, S)$ to result in an $(a, b, \tau)$ streaming code.

\eprf

\subsection{Equivalent Conditions on Parity Check (P-C) Matrix}
Motivated by the p-c-matrix-based properties for DE-based \params\ streaming codes given in \cite{NikDeepPVK}, we list down here analogous conditions for SDE-based streaming codes. These conditions will be used in proving that the linear field size code to be presented in Section~\ref{sec:constr}  is an \params\ streaming code. We first state, without proof, a well-known result that is useful in coming up with these conditions.
\blem \label{lem:pc}
Let $\mathbb{C}$ be an $[n,k]$ linear code over $\mathbb{F}_q$ and let $H\triangleq[\underline{h}_0~\underline{h}_1 ~\dots~ \underline{h}_{n-1}] \in \mathbb{F}_q^{(n-k)\times n}$ be a p-c matrix for $\mathbb{C}$, where $\underline{h}_i \in \mathbb{F}_q^{n-k}$ denotes the $i$th column of $H$. Let $E \subseteq [0:n-1] $ be an erasure set such that $i \in E$, then the code symbol $c_i$ can be recovered iff $\underline{h}_i \notin \text{span} \left \langle\{\underline{h}_j~|~ j \in E \setminus \{i\} \} \right \rangle$.
\elem

Let $\mathbb{C}$ be an $[n,k]$ linear code over $\mathbb{F}_q$ and $P \subseteq [0:n-1]$. Then the punctured code $\mathbb{C}|_{P}$ is the code of block length $|P|$ obtained by deleting all the coordinates in $[0:n-1] \setminus P$. Let $H$ be a p-c matrix for $\mathbb{C}$ and $H^{(i)}\triangleq[\underline{h}_{0}^{(i)}~\dots~ \underline{h}_{r_i}^{(i)}]$ be the p-c matrix for $\mathbb{C}|_{[0:r_i]}$, for all $i \in [0:n-1]$. Here $\underline{h}_{j}^{(i)}$ denotes $j$th column of $H^{(i)}$.   
Using Lemma \ref{lem:pc}, the recovery conditions in Theorem~\ref{thm:sde} can be restated in terms of these p-c matrices and placement set $S$ as follows:

\begin{enumerate}

	\item Random erasure recovery: for all $i \in [0:n-1]$ and  every $A \subseteq [i:r_i]$ such that $i \in A$, $|A| \le a$,

\bit 
\item if $r_i < n-1$,   $\underline{h}_i^{(i)} \notin \text{span}  \langle\{\underline{h}_j^{(i)}~|~ j \in A \setminus \{i\} \}  \rangle $, 
\item else if $r_i=n-1$, $\{\underline{h}_j~|~ j \in A \}$ is a linearly independent set. 
\eit 
\item Burst erasure recovery: for all $i \in [0:n-1]$ and every $B \subseteq [s_i:\min\{s_i+b-1,N-1\}]$ such that $s_i \in B$,
\bit 
\item if $r_i < n-1$, $\underline{h}_i^{(i)} \notin \text{span} \left <\{\underline{h}_j^{(i)}~|~ j \in f_S(B)\setminus \{i\} \} \right >,$ 
\item else if $r_i=n-1$, $\{\underline{h}_j~|~ j \in f_S(B) \}$ is a linearly independent set. 
\eit 
\item[]
\end{enumerate}

We now state a result which makes checking these p-c conditions easier in some cases. We will make use of this result repeatedly in the proof of Theorem~\ref{Thm_SDE}.   
\blem 
Let $i \in [p:r_p]$ and $T \subseteq [i+1:r_i]$. If  $\underline{h}_i^{(p)} \notin \text{span} \left <\{\underline{h}_j^{(p)}~|~ j \in  T \cap [i+1:r_p]\} \right > $ , then $\underline{h}_i^{(i)} \notin \text{span} \left <\{\underline{h}_j^{(i)}~|~ j \in T \} \right > $.
\elem  
\bprf
Note that $i \ge p \implies r_i \ge r_p$ and hence $H^{(p)}$ is a sub-matrix of $H^{(i)}$ as shown below. If $r_i=r_p$, the statement trivially holds. For $r_i > r_p$, $H^{(i)}$ has the following structure:
\bean
H^{(i)} &=& \left[ \begin{array}{cc}
H^{(p)} & 0 \\
M_1 & M_2
\end{array} \right] = \left[ \begin{array}{cccc}
\underline{h}_0^{(i)} & \underline{h}_1^{(i)} & \cdots & \underline{h}_{r_i}^{(i)}
\end{array}\right].
\eean
Suppose the statement doesn't follow then:
\bean
\underline{h_i}^{(i)} &=& \sum\limits_{j \in T} a_j \underline{h}_{j}^{(i)}.
\eean
By equating rows where columns $h_{r_p+1}^{(i)}, \cdots, h_{r_i}^{(i)}$ have zeros we have:
\bean
\underline{h_i}^{(p)} &=& \sum\limits_{j \in T \cap [i+1, r_p]} a_j \underline{h}_{j}^{(p)}.
\eean
This contradicts our assumption that $h_i^{(p)} \notin \text{span} \left <\{\underline{h}_j^{(p)}~|~ j \in  T \cap [i+1:r_p]\} \right >$.
\eprf
\section{Building Blocks} \label{sec:building}

In this section, we provide a construction of the scalar base code, which will be used in conjunction with SDE to obtain rate-optimal linear field size \params\ streaming codes.

\begin{defn}[Zero-band MDS Generator Matrix] 
\normalfont A $k\times n$ matrix $Z=(z_{ij})_{i \in [0:k-1],j\in[0:n-1]}$ is a zero-band MDS generator matrix if: 
\begin{enumerate}
	\item $Z$ is a generator matrix for an $[n,k]$ MDS code and,
	\item $z_{ij}=0$, $ \forall \{i,j\}$ such that $j \in [i+1:i+k-1] (mod \ n).$
\end{enumerate}
\edefn
Note that $z_{ij} \neq 0$, $ \forall \{i,j\}$ such that $j \notin [i+1:i+k-1] (mod \ n)$. Otherwise, it would contradict the minimum distance of MDS code being equal to $n-k+1$.
\begin{figure}[htbp]
\bean
\left[ \begin{array}{cccccc}
* & 0 & 0 & * & * & *\\
* & * & 0 & 0 & * & *\\
* & * & * & 0 & 0 & * \\
\end{array} \right]
\eean
	\caption{The structure of a $3 \times 6$ zero-band MDS generator matrix. Here * is a place-holder for non-zero field elements.}
\label{fig:zb}
\end{figure}


For every set of positive integers $\{k,n\}$ such that $n \ge k$, a $k\times n$ zero-band MDS generator matrix can be explicitly constructed over $\mathbb{F}_q$, if $q \ge n$ (for instance, see \cite{NikDeepPVK}).

\begin{defn}[Super-regular Matrix]\normalfont  A $k\times n$  matrix $C$ is a super-regular matrix if every square sub-matrix of $C$ is invertible.	
\edefn
It is a well-known result \cite{ecc_Mac_Slo} that a $k\times n$ Cauchy matrix is super-regular and can always be explicitly constructed over $\mathbb{F}_q$, with $q \ge k+n$, for all positive integers $k,n$.   

\begin{constr} \label{base}\normalfont
	Here we construct an $[n=\rho-a+r,~k=\rho-a] $ linear block code $\mathbb{C}_{a,r,\rho}$ for all $\{a,r,\rho\}$ such that $r=\ell a$, for some integer $\ell \ge 1$, and $r < \rho$. 
	Let $Z=[Z_1~Z_2]$ be an $a \times 2a$ zero-band MDS generator matrix, where $Z_1=Z(0:a-1,~0:a-1)$ and $Z_2=Z(0:a-1,~a:2a-1)$. Let $C$ be an $r \times (\rho-r)$ Cauchy matrix.
	We now describe an $r \times n$ p-c matrix $H$ of  $\mathbb{C}_{a,r,\rho}$ through a series of steps.     
	\begin{enumerate} 
		\item Initialize  $H$ to be the $r\times n$ all-zero matrix, 
		\item set $H(0:a-1,~0:a-1)=I_a$,
		\item set $H(ia:ia+a-1,~ia:ia+a-1)=Z_1, ~\forall i \in [1:\ell-1]$,
		\item set $H(0:r-1,~r:\rho-1)=C$,
		\item set $H(ia:ia+a-1,~\rho+(i-1)a:\rho+(i-1)a+a-1)=Z_2,~\forall i \in [1:\ell-1]$.
	\end{enumerate} 	  
\end{constr}
	\begin{figure}[htbp]
	\centering
	\includegraphics[width=0.7\textwidth]{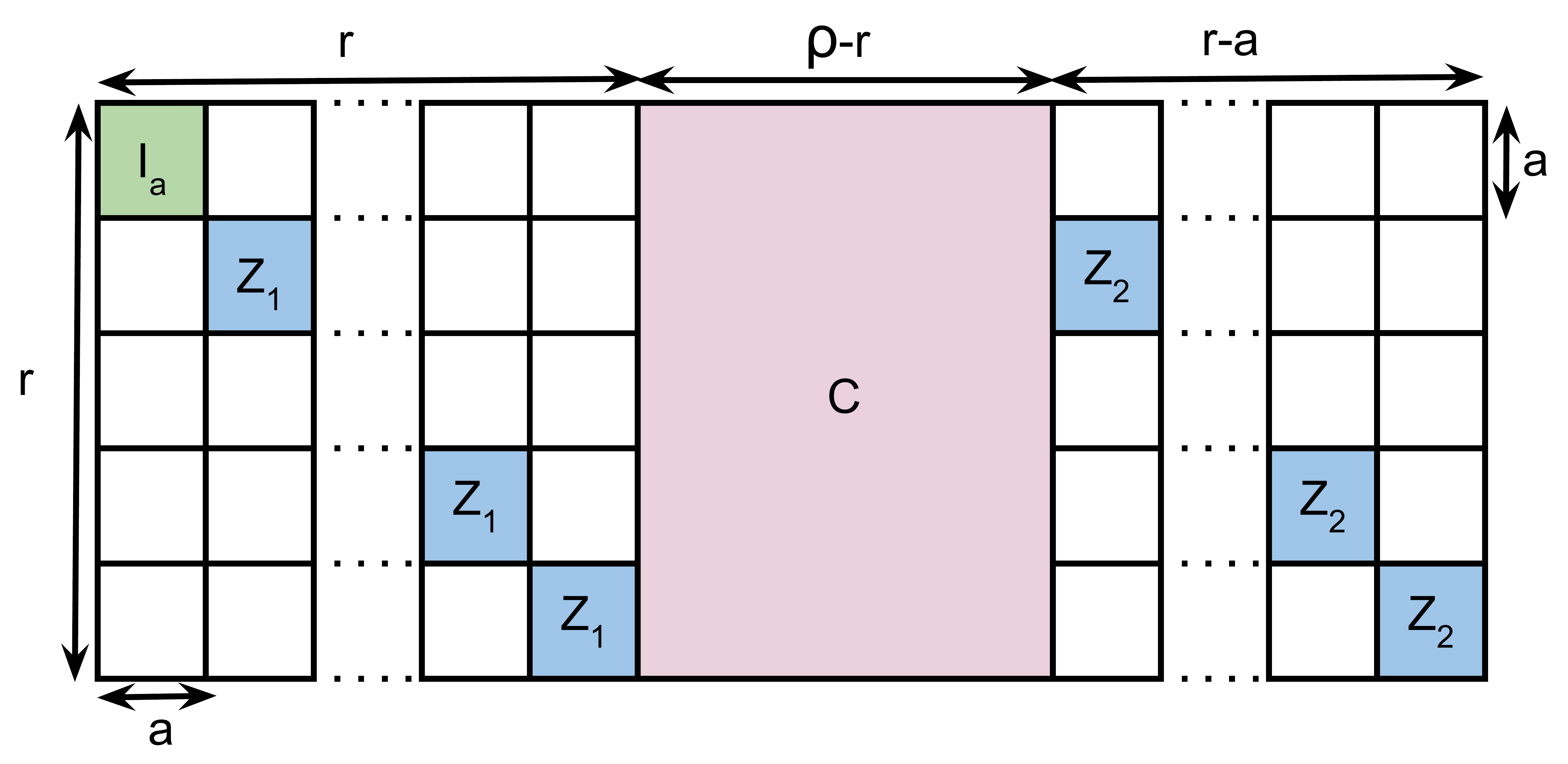}
	\caption{P-C matrix of $\mathbb{C}_{a,r,\rho}$}.
	\label{fig:base}
\end{figure} 
 
\section{Linear Field Size Construction for $gcd(b,\tau+1-a)\ge a$ } \label{sec:constr}

In this section, we present a linear field size rate-optimal \params\ streaming code construction for all $\{a,b,\tau\}$ satisfying $gcd(b,\tau+1-a)=a+g \ge a$, where $g$ is a non-negative integer.
 Clearly, there exist positive integers $\ell,m$ such that $b=\ell(a+g)$, $\tau+1-a=m(a+g)$ and $gcd(m,\ell)=1$. It also follows that $m \ge \ell$, otherwise $\tau+1=m(a+g)+a \le b$. This is not possible as $b \le \tau$.

Now, we will use  $\mathbb{C}_{a,\ell a, (m+1)a}$ (see Construction \ref{base}) as the base code. It can be easily seen that $\mathbb{C}_{a,\ell a, (m+1)a}$ is an
$[n=(m+\ell)a,~k=ma]$ code and its rate $R=\frac{m}{m+\ell}= \frac{\tau+1-a}{\tau+1-a+b}=R_\text{opt} $. We remark that both the matrices $C$ and $Z$ used in the construction exist over $\mathbb{F}_q$ if $q \ge (m+1)a$ and hence, $\mathbb{C}_{a,\ell a,(m+1)a}$ requires only a field of size $q\ge(m+1)a= \frac{a(\tau+1+g)}{a+g} = O(\tau)$. For SDE, we fix the dispersion span  $N=(m+\ell-1)(a+g)+a$ and choose  the placement set:
\bea
\label{eq:placeset}
S_{a,b,\tau}=\bigcup\limits_{i=0}^{i=m+\ell-1}[i(a+g):i(a+g)+a-1].
\eea 
It can be clearly verified that $|S_{a,b,\tau}|=(m+\ell)a=n$. 


\subsection{An Example: $\{a=2,b=6,\tau=10\}$}
Here $gcd(b,\tau+1-a)=3$. Hence, we have $m=3, \ell=2, g=1$, $S_{2,6,10}=\{0,1,3,4,6,7,9,10,12,13\}$, $N=14$ and $\mathbb{C}_{2,4,8}$ is an $[10,6]$ code.  The p-c matrix $H$ of $\mathbb{C}_{2,4,8}$ is given by:
 \bean
 H = 
 \begin{bmatrix}
 1 & 0 & 0 & 0 & c_{00} & c_{01} & c_{02} &  c_{03} & 0 & 0 \\
 0 & 1 & 0 & 0 & c_{10} & c_{11} & c_{12} &  c_{13} & 0 & 0\\
 0 & 0 & z_{00} & 0 & c_{20} & c_{21} & c_{22} &  c_{23} & z_{02} & z_{03}\\
 0 & 0 & z_{10} & z_{11} & c_{30} & c_{31} & c_{32} &  c_{33} & 0 & z_{13}
 \end{bmatrix},
 \eean 
 such that:
 \bean
 Z= \begin{bmatrix}
 	z_{00} & 0 & z_{02} & z_{03}\\
 	z_{10} & z_{11} & 0 & z_{13}
 \end{bmatrix}, C= \begin{bmatrix}
 c_{00} & c_{01} & c_{02} &  c_{03} \\
 c_{10} & c_{11} & c_{12} &  c_{13} \\
 c_{20} & c_{21} & c_{22} &  c_{23} \\
 c_{30} & c_{31} & c_{32} &  c_{33} 
\end{bmatrix}
 \eean
 are $2 \times 4$ zero-band MDS generator matrix, $4 \times 4$ Cauchy matrix respectively.
Both these matrices can be constructed over $\mathbb{F}_8$. 
 In order to prove that the packet-level code constructed by SDE of $\mathbb{C}_{2,4,8}$ with placement set $S_{2,6,10}$ is a rate-optimal streaming code, we only have to  show that  $\mathbb{C}_{2,4,8}$ along with  $S_{2,6,10}$ satisfy both random and burst erasure recovery conditions laid out in Section \ref{sec:sde}. 
 In this case, we have: $r_0=7, r_1=7, r_i=9, \forall i \in [2:9]$.
 
 \subsection*{Random Erasure Recovery} 
  \bit 
 \item  $ r_i < 9 \implies i\in \{0,1\}$ 
 \eit  
  The p-c matrix of punctured code $\mathbb{C}_{2,4,8}|_{[0:7]}$  takes the form    
 	  $\begin{bmatrix}
 	 	1 & 0 & c_{00} & c_{01} & c_{02} &  c_{03}  \\
 	 	0 & 1 & c_{10} & c_{11} & c_{12} &  c_{13} 
 	 \end{bmatrix}$, after removing columns $2, 3$ that are all-zero.  This matrix can be shown to be a generator matrix for a $[6,2]$ MDS code and hence no two columns are linearly dependent. Thus, random erasure recovery condition is satisfied for $i=0,1$.
 	  \bit 
 	 \item  $r_i=9 \implies i \in [2:9]$
 	 \eit  
 Here we need to show that any two columns among last $8$ columns of $H$  are linearly independent. It can be easily seen that no two among columns $2,3,8$ and $9$ can be linearly dependent as $Z$ is a generator matrix for a $[4,2]$ MDS code. Since every square sub-matrix of a Cauchy matrix is invertible, any two among columns $4,5,6$ and $7$ are linearly independent. Also, if we pick one column with index in $\{2,3,8,9\}$ and another column with index in  $\{4,5,6,7\}$, they are linearly independent because they have different support. Thus we have showed that no two columns of $H(0:3,~2:9)$ are linearly dependent, thereby showing that random erasure condition is satisfied for $i \in [2:9]$.  

  \subsection*{Burst Erasure Recovery} 
For $j\in[0:8]$, consider any consecutive $b=6$ columns in $[0:13]$ of the form $B\triangleq[j:j+5]$. The placement set $S_{2,6,10}$ ensures that $|f_{S_{2,6,10}}(B)|=4$. Hence it follows that any burst of size $6$ results in a loss of $4$ consecutive symbols for every underlying codeword of the base code (for instance, see Fig.~\ref{fig:sde}).


 \bit 
 \item  $ r_i < 9 \implies i\in \{0,1\}$ 
 \eit  
In $H(0:1,0:7)$, which is the p-c matrix of $\mathbb{C}_{2,6,10}|_{[0:7]}$, it can be easily seen that column $0$ is not a linear combination columns $1,2$ and $3$ due to disjoint support. Similarly, column $1$ of $H(0:1,0:7)$ does not lie in span of columns $2,3$ and $4$, due to difference in support. Thus, for $i=0,1$ burst erasure recovery condition is satisfied.   
\bit 
 \item $r_i=9 \implies i \in [2:9]$
 \eit 
 In order to show that burst erasure property holds for $i=[2:9]$, it suffices to prove that any collection of $4$ consecutive columns among last $8$ columns of $H$ forms a linear independent set.  Equivalently, one needs to show that $H(0:3,i:i+3)$ is invertible, for all $i \in [2:6]$. As $\begin{bmatrix}
 z_{00} & 0 \\
 z_{10} & z_{11}
 \end{bmatrix}$ and $\begin{bmatrix}
 c_{00} & c_{01} \\
 c_{10} & c_{11}
 \end{bmatrix}$ are both invertible, $H(0:3,~2:5)$ is invertible. As $\begin{bmatrix}
  c_{00} & c_{01} & c_{02}\\
  c_{10} & c_{11} & c_{12} \\
   c_{20} & c_{21} & c_{22}
  \end{bmatrix}$ is invertible and $z_{11} \neq 0$, invertibility of $H(0:3,~3:6)$ follows. The matrix $H(0:3,~4:7)$ is the Cauchy matrix $C$ and is hence invertible. The invertibility of 
     $\begin{bmatrix}
 c_{01} & c_{02} & c_{03}\\
 c_{11} & c_{12} & c_{13} \\
 c_{31} & c_{32} & c_{33}
 \end{bmatrix}$ together with $z_{02} \neq 0$ results in $H(0:3,~5:8)$ being invertible. Since $\begin{bmatrix}
 c_{02} & c_{03} \\
 c_{12} & c_{13}
 \end{bmatrix}$ and $\begin{bmatrix}
 z_{02} & z_{03} \\
 0 & z_{13}
 \end{bmatrix}$ are both invertible, $H(0:3,~6:9)$ is invertible. Thus, we have proved that the packet-level code SDE$(\mathbb{C}_{2,4,8},S_{2,6,10})$ is a $(a=2,b=6,\tau=10)$ rate-optimal streaming code and can be constructed over $\mathbb{F}_8$.  This example is generalized in the Theorem stated below.
 \bthm \label{Thm_SDE}
 
 For any set of parameters $\{a,b,\tau\}$ such that $a+g=gcd(b,\tau+1-a) \ge a$, let $\ell=\frac{b}{a+g}$ and $m=\frac{\tau+1-a}{a+g}$, then the packet level code SDE$(\mathbb{C}_{a,\ell a,(m+1)a}, S_{a,b,\tau})$, is an $(a,b,\tau)$ streaming code that is rate-optimal and $O(\tau)$ field size.
 \ethm 
 \bprf 
  When gcd$(b,\tau+1-a)=b$, we have $\ell=1$ and $\mathbb{C}_{a,a ,(m+1)a}$ is an $[(m+1)a,a]$ MDS code. The packet level code for $\ell=1$ case is exactly same as the MDS base code construction in \cite{simple}. We skip the proof for $\ell=1$ case  since it is provided in \cite{simple}. Throughout the reminder of the proof we assume $\ell >1$.\\ 
  
 Let $H$ denote the $\ell a \times (m+\ell)a$ p-c matrix of $\mathbb{C}_{a,\ell a ,(m+1)a}$ and $S_{a,b,\tau} = \{s_0, \cdots, s_{n-1}\}$ is as defined as shown in equation \eqref{eq:placeset}. Therefore $N-1=s_{n-1}=(m+\ell-1)(a+g)+a-1$. The index of further most symbol accessible for recovery of symbol at index $ja$ is given by:
 \bean
 r_{ja} &=& f_{S_{a,b,\tau}}(\min\{s_{ja}+\tau, N-1\})\\
        &=& f_{S_{a,b,\tau}}(\min\{(j+m)(a+g)+a-1, N-1\})\\
        &=& \begin{cases}
        	(j+m+1)a-1 & j < \ell \\
        	(m+\ell)a-1 & j \ge  \ell.
        \end{cases}
 \eean
 The values of $r_{ja}$ determines the punctured codes which we need to consider to prove the theorem.\\ 
 
 We will now show that burst $B$ in $[0, N-1]$ of size $b$ results in burst of size $\ell a $ in $[0, n-1]$. Let $i = ja+j' \in [0:n-1]$ where $j' \in [0:a-1]$ and $B = [s_i:s_i+b-1] \subseteq [0: N-1]$ then:
 \bean
 f_{S_{a,b,\tau}}(B) &=& f_{S_{a,b,\tau}}(\{j(a+g)+j_0 \mid j_0 \in [j': j'+\ell(a+g)-1] \})\\
 &=& f_{S_{a,b,\tau}}(\{j(a+g)+j_0 \mid j_0 \in [j': a-1] \}) \cup\\ 
 && \cup_{j_1=j+1}^{j+\ell-1}f_{S_{a,b,\tau}}(\{j_1(a+g)+j_0 \mid j_0 \in [0: a-1] \}) \cup \\
&& f_{S_{a,b,\tau}}(\{(j+\ell)(a+g)+j_0 \mid j_0 \in [0: j'-1] \})\\
&=&  [i: (j+1)a-1] \cup \left( \cup_{j_1=j+1}^{j+\ell-1} [j_1a: (j_1+1)a-1]  \right)\\ && \ \ \cup [(j+\ell)a: (j+\ell)a+j'-1] \\
&=& [i: i+\ell a -1].
 \eean
 
 Therefore, by Theorem~\ref{thm:sde} and the fact that $f_{S_{a,b,\tau}}(B) = [i: i+\ell a -1]$, it is enough to show that for all $i \in [0:n-1]$, $c_i$ can be recovered from $\{ c_j \mid j < i\} \cup \{c_j \mid j \in [i+\ell a: r_i]\}$ to prove burst erasure recovery.\\
  
  We divide the proof into three parts based on the value of $i$. 
 
  \ben
  \item[] 
  \item For $i \in [0:a-1]$, we show how to recover symbol $i$ by accessing symbols only until $r_0=(m+1)a-1$, though we have access until $r_i$-th symbol. The parity check matrix that represents the punctured code $\mathbb{C}_{a, \ell a, (m+1)a} \mid_{r_0}$ is given by:
  \bean
  H^{(0)}=\left[\begin{array}{ccc}
   I_a & 0_{a \times (\ell-1)a} & C(0:a-1, 0:(m+1-\ell)a-1) 
  \end{array}\right]
  \eean
  \emph{Random Erasure Recovery:} Let $A \subseteq [i:r_i]$ be a set of erasures such that $|A| \le a$. Let $A_0 = [i:r_0] \cap A$. Then it is clear that $|A_0| \le a$. Clearly the $i$-th column in $H^{(0)}$, $h_i^{(0)}$ doesn't belong to span of any $a-1$ other columns in $[i+1, r_0]$ as $C$ is a Cauchy matrix. \\
 
  \emph{Burst Erasure Recovery:} Let $B=[s_i, s_i+b-1]$. This will result in erasures $f_{S_{a,b,\tau}}=[i,i+\ell a-1]$ in the base code. We will show that the code symbol $c_i$ can be recovered by accessing symbols until $r_0$. Here, $h_i^{(0)}$ is clearly not in span of $\{h_{i+1}^{(0)}, \cdots, h_{a-1}^{(0)}, h_{\ell a}^{(0)}, \cdots, h_{\ell a + i}^{(0)} \}$, again due to super-regular property of $C$, and hence $c_i$ can be recovered.

  \item[] 
  \item For $i \in [ja, (j+1)a-1]$ with $j \in [1:\ell-2]$ we show how to recover $c_i$ by accessing symbols only until $r_{ja}=(j+m+1)a-1$ though we have access until $r_i$. The parity check matrix that represents the punctured code $\mathbb{C}_{a,\ell a, (m+1)a} \mid _{r_{ja}}$, $H^{(ja)}$ is given by:
  \bean
H^{(ja)} = \left[\begin{array}{c|c|c|c|c|c|c|c|c}
  	I_a & & & & \multirow{4}{*}{$0$} &  \multirow{4}{*}{$C(0:(j+1)a-1, 0:(m+1-\ell)a-1)$} &\\ \cline{1-4}\cline{7-9}
  	& Z_1 & & & & & Z_2 & &\\ \cline{1-4}\cline{7-9}
  	& & \ddots & & & & &\ddots\\ \cline{1-4}\cline{7-9}
  	& & & Z_1 & & & & & Z_2
  	\end{array}\right]
  \eean
  It can observed that columns of $H^{(ja)}$ with index in $[(j+1)a: \ell a-1]$ are all-zero columns and hence erasures in these columns can be neglected. \\
  
  \emph{Random Erasure Recovery}:  Let $A \subseteq  [i:r_{ja}] \setminus [(j+1)a:\ell a-1] $ with $|A| \le a$ and $i \in A$ be such that $\underline{h}_i^{(ja)} = \sum_{p \in A \setminus \{i\}} u_p\underline{h}_p^{(ja)}$, for some assignment of $u_p \in \mathbb{F}_q$. In $\underline{h}_i^{(ja)}$ the first $a$ rows are zeros and columns with index in $[\ell a:(m+1)a-1]$ are the only columns with non-zero entries in first $a$ rows. If one column from $[\ell a:(m+1)a-1]$ is involved in the linear combination, then $a$ other columns from $[\ell a:(m+1)a-1]$ are required to obtain zeros in first $a$ rows, because of the super-regular property. Hence, we have $u_p = 0$ for all $p \in A \cap [\ell a:(m+1)a-1]$. Also, $\underline{h}_i^{(ja)}$ has zeros in rows $[a:ja-1]$. It can be seen that no collection of $\le a-1$ columns from $[(m+1)a: r_{ja}-a] $, can linearly combine to form zeros in  rows $[a:ja-1]$, because of the support structure of columns and MDS property of $Z$ implying that $u_p = 0$ for all $p \in A \cap [\ell a, r_{ja}-a]$. The remaining columns $A \cap \left([i+1, (j+1)a-1] \cup [r_{ja}-a+1, r_{ja}]\right)$ can not span $i$-th column, again due to MDS property of $Z$. Therefore it is not possible to have: $h_i^{(ja)} = \sum_{p \in A \setminus \{i\}} u_p\underline{h}_p^{(ja)}$. Hence $c_i$ is recoverable from $[i:r_{ja}] \setminus A$ for any $A \subseteq [i:r_{ja}]$ such that $i \in A$ and $|A| \le a$.\\
  
  \emph{Burst Erasure Recovery:} Let $B = [s_i:s_i+b-1]$, then the base code sees erasures $B_0 = [i:i+\ell a-1]$. We want to show that $\underline{h}_i^{(ja)}$ doesn't belong to span of columns of $H^{(ja)}$ indexed by elements in $B_0 \setminus \{i\}$. It is enough to consider columns in $B_0 \setminus [(j+1)a:\ell a -1]$ as the columns $[(j+1)a:\ell a -1]$ are all zero.  
  
  \ben
  \item[] 
  \item[(a):] For $i\le (m-\ell+1)a$. The submatrix formed by columns $B_0 \setminus [(j+1)a:\ell a-1]$ in matrix $H^{(ja)}$ is of the form:
  \bean
  \hat{H}^{(ja)}=\left[\begin{array}{c|c}
  	0 & C(0:i-1,0:i-1)\\ \hline
   Z_1(i-ja:a-1, i-ja:a-1) & C(i:(j+1)a-1,0:i-1)
  \end{array}\right].
  \eean
  Note that the number of rows of this matrix is same as the rows of $H^{(ja)}$ which is $(j+1)a$ and the number of columns is given by $\ell a - (\ell a - (j+1)a) = (j+1)a$. Clearly this matrix is invertible as both  $C(0:i-1,0:i-1)$ and $Z_1(i-ja:a-1; i-ja:a-1)$ are invertible. The invertibility of $Z_1(i-ja:a-1; i-ja:a-1)$ can be easily argued using MDS property of Z and lower triangular structure of $Z_1$.
  \item[] 
  \item[(b):] For $i > (m-\ell+1)a$. Let $i = ja+j'$ for some $j' \in [0, a-1]$, then $i+\ell a-1 = (j +\ell)a+j'-1$. Here $j_1 = (j+\ell-m-1)$ is the number of $Z_2$ blocks that appear in the submatrix formed by columns $B_0 \setminus [(j+1)a:\ell a-1]$ of $H^{(ja)}$. The form of this submatrix is given by:
  \bean
 \hat{H}^{(ja)}=\left[\begin{array}{c|c|c|c|c|c}
  	0 & \multirow{6}{*}{$C^*$} & & & &\\ \cline{1-1} \cline{3-6}
  	0 &  & Z_2 & & &\\ \cline{1-1} \cline{3-6}
  	\vdots &  & & \ddots & &\\ \cline{1-1} \cline{3-6}
  	 & & & & Z_2 &\\ \cline{1-1} \cline{3-6}
  	 & & & & & \underbrace{Z_2^*}_{(i-ja) \times (i-ja)}\\
   	 & & & & & \underbrace{0}_{((j+1)a-i) \times (i-ja)}\\ \cline{1-1} \cline{3-6}
  	  & & & & & 
  	 \\ \cline{1-1} \cline{3-6}
  	 \underbrace{0}_{(i-ja) \times ((j+1)a-i)} & & & & & \\
  	 \underbrace{Z_1^*}_{((j+1)a-i) \times ((j+1)a-i)} & & & & & 
  	\end{array}\right].
  \eean
  where $Z_1^*=Z_1(i-ja:a-1, i-ja:a-1)$, $Z_2^*=Z_2(0:i-ja-1,0:i-ja-1)$ and $C^*=C(0:(j+1)a-1, 0:(m+1-\ell)a-1)$.
  Consider the rows of  $\hat{H}^{(ja)}$ with non-zero support only in Cauchy columns. These $(m-\ell+1)a$ rows are indexed by $[0,a-1] \cup  [ i-(m-\ell)a: i-1]$. The submatrix of $C$, denoted by $\hat{C}$, containing only these $(m-\ell+1)a$ rows is square and hence invertible.
  The lower triangular structure of $Z_1$ and upper triangular structure of $Z_2$ along with MDS property of $Z$ gives invertibility of $Z_1^*$ and $Z_2^*$ respectively. Note that $\hat{H}(ja)$ contains $a$ columns from $Z_1^*$ and $Z_2^*$ together and from the structure of $Z$ it follows that there is row in  $ \hat{H}^{(ja)}$ with non-zero entry from both $Z_1^*$ and $Z_2^*$.  By row and column permutation, the  non-Cauchy columns of $\hat{H}$ can be made to a block diagonal matrix, with each block invertible. Using this and invertibility of $\hat{C}$, it can be inferred that the matrix $ \hat{H}^{(ja)}$ is invertible.
  \item[] 
  \een 
  Thus, we have proved that the submatrix formed by columns $B_0 \setminus [(j+1)a:\ell a-1]$ in matrix $H^{(ja)}$ is is invertible and hence $\underline{h}_i^{(ja)}$ doesn't lie in span of columns of $H^{(ja)}$ indexed by $B_0 \setminus \{i\}$.
  \item[] 
  \item For $i \in [(\ell-1)a:(m+\ell)a-1]$, the value of $r_i = (m+\ell)a-1$.\\
    
    \bean
  H(0:r-1,~(\ell-1)a:(m+\ell)a-1)  = \left[\begin{array}{c|c|c|c|c}
  & \multirow{4}{*}{$C_{r \times (m+1-\ell)a}$} & & \\
  \cline{1-1} \cline{3-5} & & Z_2 & & \\
  \cline{1-1} \cline{3-5} & & & \ddots \\  \cline{1-1} \cline{3-5}
  Z_1 &  &  & & Z_2
  \end{array}\right]
  \eean
  
  \emph{Random Erasure Recovery:} It is to be shown that any collection of $a$ columns of $H$ with index in  $[(\ell-1)a:(m+\ell)a-1]$ forms a linearly independent set. 
   Suppose there exists $A \subseteq  [(\ell-1)a:(m+\ell)a-1] $ with $|A| \le a$ such that $\sum_{j \in A} u_j\underline{h}_j=0$, $u_j \in \mathbb{F}_q$. Since columns with index in $[\ell a:(m+1)a-1]$ are the only columns with non-zero entries in first $a$ rows, if one column from $[\ell a:(m+1)a-1]$ is part of the linear combination, then $a$ other columns from $[\ell a:(m+1)a-1]$ are required to obtain zeros in first $a$ rows, because of the super-regular property. Hence, $u_j = 0$ for all $j \in A \cap [\ell a:(m+1)a-1]$. Now, because of the MDS property of 
  $Z$ and support structure of columns, no collection of $a$ columns with index in $[(m+1)a:(m+\ell)a-1] \cup [(\ell-1)a:\ell a-1]$ can linearly dependent. Thus we have $u_j = 0$ for all $j \in A$. Therefore all the columns indexed by elements in $A$ are linearly independent and hence random erasure recovery is guaranteed.\\
  
  \emph{Burst Erasure Recovery:} To prove this property, it suffices to show that the square sub-matrix formed any $r=\ell a$ consecutive columns of $H(0:r-1,~(\ell-1)a:(m+\ell)a-1)$ is invertible. Consider some set $B_0$ consisting of r consecutive integers from $[(\ell-1)a:(m+\ell)a-1]$.  Let $\hat{H}$ be the submatrix of  $H(0:r-1,~(\ell-1)a:(m+\ell)a-1)$ which is formed by collecting columns indexed by $B_0$.  
  
  It can be verified that $\hat{H}$ has $\big|B_0 \cap [\ell a:(m+1)a-1]\big|$ rows which has support only in the Cauchy part and the sub-matrix of $C$ formed by these rows is a square and invertible. By row and column permutation, the  non-Cauchy columns of $\hat{H}$  (if they exist) can be made to a block diagonal matrix. Since $m-\ell+1 \ge 1$, less than $a$ columns with index in $[(\ell-1)a:\ell a-1] \cup [(m+\ell-1)a:(m+\ell)a-1]$ are part of  $\hat{H}$. This ensures that no  more than $a$ consecutive columns of same $Z$ matrix are involved in $\hat{H}$. Now, using this fact and properties of $Z$, it can easily argued that each of these block are invertible, thus proving invertibility of  $\hat{H}$.
  
  \een

 \eprf

\section{Diagonal Embedding Based Constructions}  \label{sec:de_constr} 

In the streaming code construction presented in Section~\ref{sec:sde}, when $gcd(b,\tau+1-a)=a$, we have $N=n$ and SDE reduces to DE. The DE of same scalar code given by Construction~\ref{base} results in a streaming code even when $gcd(b,\tau+1-a) < a$ as long as $a | b, \tau+1-a \ge b$. We also come up with a modified scalar code shown in construction \ref{constr1} whose DE results in streaming codes when $b \mod a = a-1$ and $\tau+1-a \ge b$.

\subsection{$a|b$ and $\tau+1-a \ge b$}

Let $\{a,b,\tau\}$ be such that $\tau+1-a \ge b$ and $b=\ell a$, where $\ell \ge 1$ is a positive integer. Then, DE of $\mathbb{C}_{a,b,\tau+1}$ results in  rate-optimal $(a,b,\tau)$ streaming code over a finite field of size $q \ge  \tau+1 $. We remark that rate of $\mathbb{C}_{a,b,\tau+1}$, $R=\frac{\tau+1-a}{\tau+1-a+b}=R_\text{opt}$. The example shown in previous section $\mathbb{C}_{2,4,8}$ results in $(a=2, b=4, \tau=7)$ streaming code by DE. Similarly, DE of  $\mathbb{C}_{2,4,7}$ results in an $(a=2, b=4, \tau=6)$ streaming code. With respect to the parameters $\{a=2, b=4, \tau=6\}$, note that we cannot invoke the construction in Section \ref{sec:constr} as $gcd(b,\tau+1-a)=1 < a$.

 \bthm \label{thm:de1}
For any set of parameters $\{a,b,\tau\}$ such that $\tau+1-a \ge b$ and $a|b$ , the DE of $\mathbb{C}_{a,b,\tau+1}$ gives a  rate-optimal $O(\tau)$ field size $(a, b, \tau)$ streaming code.
\ethm 
\bprf 
Random erasure recovery proof follows along the same line of proof for Theorem \ref{thm:sde}. The restriction $\tau+1 \ge b+a$ ensures that no more than $a$ coordinates associated with same $Z$ matrix are involved in the same burst erasure. Under this restriction all the arguments in burst erasure recovery proof of Theorem \ref{thm:sde} follows here as well.
\eprf

\subsection{$b \mod a=a-1$ and $\tau+1-a \ge b$}
Assume that $\{a,b,\tau\}$ is such that $\tau+1-a \ge b$ and $b=\ell a + a-1$, where $\ell \ge 1$ is a positive integer. Then, we come up with an $[n=\tau+1-a+b, ~k=\tau+1-a]$ linear code $\mathbb{C}^{I}_{a,b,\tau+1}$ over a finite field of size $q \ge \tau+1$ and using DE we obtain a rate-optimal $(a,b,\tau)$ streaming code. Note that rate of $\mathbb{C}^{I}_{a,b,\tau+1}$, $R=\frac{\tau+1-a}{\tau+1-a+b}=R_\text{opt}$. 
\begin{constr}\label{constr1} Let $Z=[Z_1~Z_2]$ be an $a \times 2a$ zero-band MDS generator matrix, where $Z_1=Z(0:a-1,~0:a-1)$ and $Z_2=Z(0:a-1,~a:2a-1)$, and $C$ be a $b \times (\tau+1-b)$ Cauchy matrix. Also, we define $Z_2^*=Z_2(0:a-2,~0:a-2)$.
 The $b \times n$ p-c matrix $H$ of $\mathbb{C}^{I}_{a,b,\tau+1}$ is given by following steps.     
\begin{enumerate} 
	\item Initialize  $H$ to be the $b\times n$ all-zero matrix,
	\item $H(0:a-2,~0:a-2)=I_{a-1}$,
	\item $H(ia-1:ia+a-2,~ia-1:ia+a-2)=Z_1,~\forall i \in [1:\ell]$,
	\item $H(0:b-1,~b:\tau)=C$,
	\item $H(a-1:2a-3,~\tau+1:\tau+a-1)=Z_2^*$,
	\item $H(ia-1:(i+1)a-2,~\tau+(i-1)a:\tau+ia-1)=Z_2, ~\forall i \in [2:\ell]$.
	
\end{enumerate}
\end{constr}
 
\begin{figure}[htbp]
	\centering
	\includegraphics[width=0.7\textwidth]{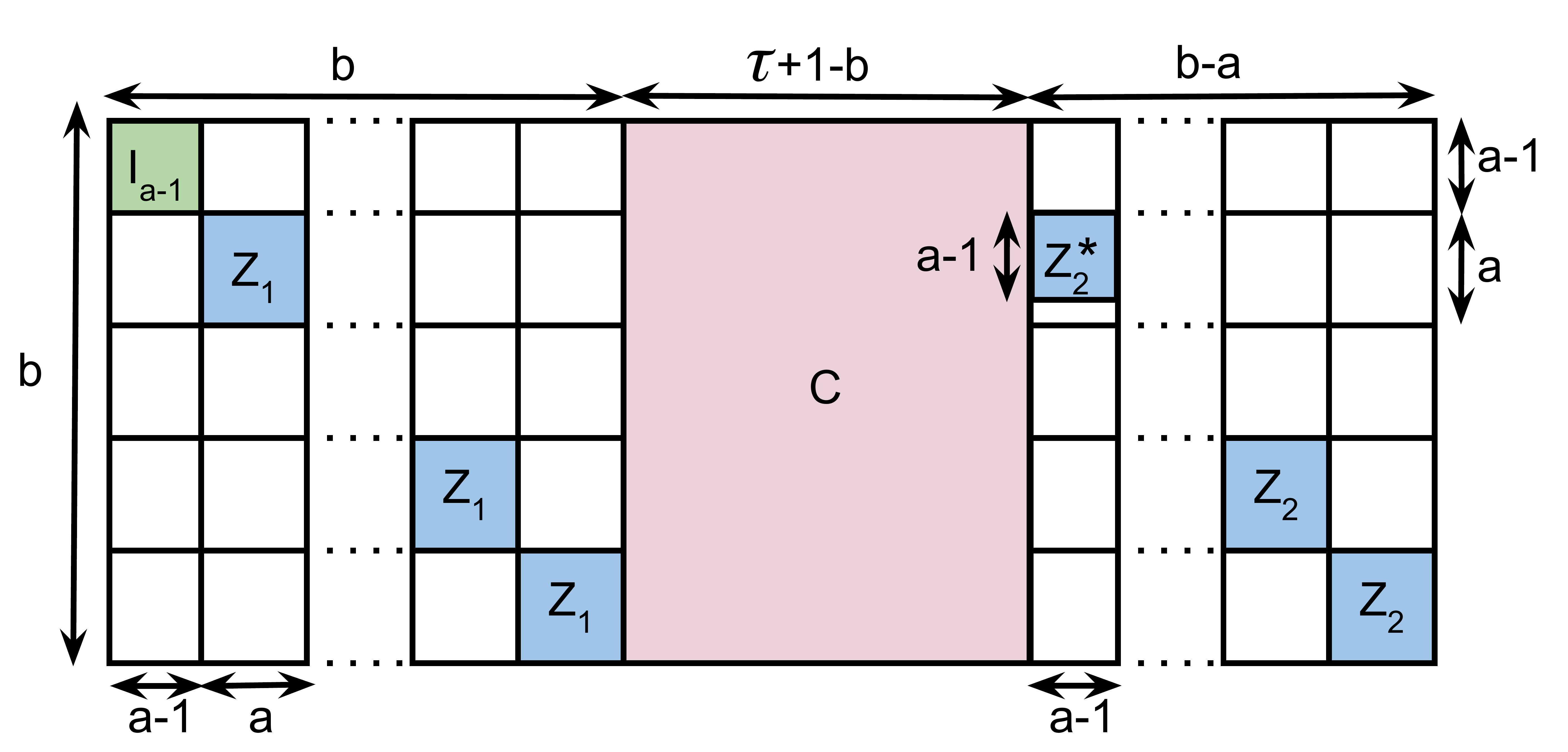}
	\caption{P-C matrix of $\mathbb{C}^{I}_{a,b,\tau+1}$}
	\label{fig:de2}
\end{figure} 

\subsubsection{An example}$\{a=2,b=5,\tau=7\}$ \\

The $\mathbb{C}^{I}_{2,5,8}$ is a $[n=11,~k=6]$ code with p-c matrix,

\begin{equation} \nonumber 
\resizebox{.5 \textwidth}{!} 
{
	$ H = 
	\begin{bmatrix}
	1 & 0 & 0 & 0 & 0 &c_{00} & c_{01} & c_{02}  & 0 & 0 & 0 \\
	0 & z_{00} & 0 & 0 & 0 & c_{10} & c_{11} & c_{12}  & z_{02} & 0 & 0\\
	0 & z_{10} & z_{11} & 0 & 0 & c_{20} & c_{21}  & c_{22} & 0 & 0 & 0\\
	0 & 0 & 0 & z_{00} & 0 & c_{30} & c_{31} & c_{32}  & 0 & z_{02} & z_{03}\\
	0 & 0 & 0 & z_{10} & z_{11} & c_{40} & c_{41} & c_{42}  & 0 & 0 & z_{13}
	\end{bmatrix}, $
}
\end{equation}
where $Z= \begin{bmatrix}
z_{00} & 0 & z_{02} & z_{03}\\
z_{10} & z_{11} & 0 & z_{13}
\end{bmatrix}$ is a $2 \times 4$ zero-band MDS generator matrix and $C=\begin{bmatrix}c_{00} & c_{10} & c_{20} & c_{30} & c_{40} & c_{5,0} \\
c_{01} & c_{11} & c_{21} & c_{31} & c_{41} & c_{5,1}\\
c_{02} & c_{12} & c_{22} & c_{32} & c_{42} & c_{5,2}  \end{bmatrix}^T$ is a $5 \times 3$ Cauchy matrix.

\bthm \label{thm:de2}
For any set of parameters $\{a,b,\tau\}$ such that $\tau+1-a \ge b$ and $b \mod a = a-1$ , the DE of $\mathbb{C}^{I}_{a,b,\tau+1}$ results in a  rate-optimal $(a, b, \tau)$ streaming code of field size $O(\tau)$.
\ethm 
\bprf Let $b = \ell a  + a -1$. 

\begin{enumerate}
	\item For $i \in [0:a-2]$, consider the p-c matrix $H^{(0)}=H([0:a-2] \cup \{2a-2\},~0:\tau)$. The columns of $H^{(0)}$ with index in $[2a-1: \ell a+a-2]$ are all-zero. The submatrix is of the form shown below:
\bean
H^{(0)} = \left[\begin{array}{c|c|c|c}
     I_{a-1} &  & \multirow{2}{*}{$0_{ a \times (\ell-1)a}$}& C(0:a-2, 0:\tau-b)\\ \cline{1-2} \cline{4-4}
     & Z_1(a-1, 0:a-1) & & C(2a-2, 0:\tau-b) 
	\end{array}\right]
\eean
\\

\emph{Random Erasure Recovery:} Fix some $i \in [0:a-2]$ and let $A \subseteq [i:\tau] \setminus [2a-1: \ell a+a-2]$ with $|A|\le a$ such that $\underline{e}_i=\underline{h}_i^{(0)} = \sum_{j \in A \setminus \{i\}} u_j\underline{h}_j^{(0)}$, $u_j \in \mathbb{F}_q $. It can be easily seen that $u_j = 0$ for all $j \in A \cap ([i+1:a-2] \cup [b: \tau])$ is necessary, as $\le a-1$ columns from  $([i+1:a-2] \cup [b: \tau])$ can not combine to form $\underline{e}_i$, because of MDS property. The remaining non-zero columns $[a-1:2a-2]$ of $H^{(0)}$ has support only in the last row and therefore cannot generate $\underline{e}_i$. Therefore it is possible to recover $c_i$ from erasure given by $A$. 
\\

\emph{Burst Erasure Recovery:} Again for a fixed $i \in [0:a-2]$, let $B=[i:i+b-1]$. Since $\tau+1-b \ge a$, we have $\max(B) < \tau$. It can be also seen that $|B \cap ([i+1:a-2] \cup [b:\tau])|=(a-2-i)+i=a-2$ and $a-2$  columns with index in $[i+1:a-2] \cup [b:\tau]$ can not linearly combine to form first $a-1$ rows of $\underline{e}_i$. Hence, burst erasure recovery condition is  satisfied for $i \in [0:a-2]$.

\item[]  
 \item For $i \in [ja-1:(j+1)a-2]$, where $j \in [1, \ell -2]$, consider the p-c matrix  $H^{(ja-1)}=H([0:(j+1)a-2] ,~0:\tau+ja-1)$. The columns of $H^{(j+a-1)}$ with index in $[(j+1)a-1: (\ell+1)a-2]$ are all-zero. 
   \bean
 H^{(ja-1)} = \left[\begin{array}{c|c|c|c|c|c|c|c|c}
 	I_{a-1} & & & & \multirow{4}{*}{$0$} &  \multirow{4}{*}{$C(0:(j+2)a-2, 0:\tau-b)$} &\\ \cline{1-4}\cline{7-9}
 	& \multirow{2}{*}{$Z_1$} & & & & & Z_2^* & &\\ \cline{7-7}
 	&  & & & & & 0 & &\\ \cline{1-4}\cline{7-9}
 	& & \ddots & & & & &\ddots\\ \cline{1-4}\cline{7-9}
 	& & & Z_1 & & & & & Z_2
 \end{array}\right]
 \eean
 
 \emph{Random Erasure Recovery}:  Let $A \subseteq  [i:i+\tau] \setminus [ja-1: (\ell+1)a-2] $ with $|A| \le a$ and $i \in A$ be such that $\underline{h}_i^{(ja-1)} = \sum_{p \in A \setminus \{i\}} u_p\underline{h}_p^{(ja-1)}$, for some $u_p \in \mathbb{F}_q$. In $\underline{h}_i^{(ja-1)}$ the first $a-1$ rows are zeros and columns $[b:\tau]$ are the only columns in $A$ with non-zero entries in first $a-1$ rows. Due to super-regular property of $C$, if one column from $[b:\tau]$ is part of  the linear combination, then at least $a$ columns from $[b:\tau]$ are needed to obtain zeros in first $a-1$ rows. Hence, we have $u_p = 0$ for all $p \in A \cap [b:\tau]$. Now, because to MDS property of $Z$, no collection of $a-1$ columns from $[i+1:(j+1)a-2]\cup[\tau+1:\tau+i]$ can span $\underline{h}_i^{(ja-1)}$. Hence, no such $A$ exists and this proves random erasure recovery for $c_i$. \\ 
 
 \emph{Burst Erasure Recovery}: Now consider a burst erasure of length $\le b$ starting at $i$, indexed by set $B=[i:i+b-1]$. Note that $|B \cap [b:\tau+b-a] = i$. If $i \le \tau+1-b$, then the square submatrix formed by non-zero columns $H^{(ja-1)}$ with index in $B$ is given by:
 \bean
 \hat{H} =\left[\begin{array}{c|c}
 	0 & C(0:i-1,0:i-1)\\ \hline
 	Z_1(i-ja+1:a-1, i-ja+1:a-1) & C(i:(j+1)a-2,0:i-1)
 \end{array}\right],
 \eean
 which is clearly invertible because $C(0:i-1,0:i-1)$ and $Z_1(i-:ja-1,i-ja+1:a-1)$ are non-singular.
 Now suppose $i > \tau+1-b$. The square submatrix formed by non-zero columns $H^{((j+1)a-1)}$ with index in $B$ has the form:
  \bean
\hat{H}^{(ja-1)}=\left[\begin{array}{c|c|c|c|c|c}
	 & \multirow{6}{*}{$C^*$} & & & &\\ \cline{1-1} \cline{3-6}
	 &  & Z_2^* & & &\\ 
	 &  & 0 &  & &\\
	 \cline{1-1} \cline{3-6}
	 &  & & \ddots & &\\
	  \cline{1-1} \cline{3-6}
	& & & & Z_2 &\\ \cline{1-1} \cline{3-6}
	& & & & & Z_2^{'}\\
	\cline{1-1} \cline{3-6}
	& & & & & \\ \cline{1-1} \cline{3-6} 
	Z_1^{'} & & & & & 
\end{array}\right]
\eean
where $C^*=C(0:(j+1)a-2, 0:\tau-b)$, $Z_1^{'}$ is a invertible submatrix of $Z_1$ and $Z_2^{'}$ is a invertible submatrix of $Z_2$. It can be shown that the number of rows of $\hat{H}^{(ja-1)}$ with support only in Cauchy part is $\tau-b$ and the submatrix of $C$ with only these rows is invertible. The condition $\tau+1-b \ge a$ ensures that no row contains non-zero entries from both $Z_1^{'}$ and $Z_2^{'}$. The invertibility of  $\hat{H}^{(ja-1)}$ follows. 

\item[] 

\item For $i \in [(\ell-1)a+a-1:\tau+b-a]$, the entire codeword is available for recovery of $c_i$.
\bean
H(0:b-1,~(\ell-1)a+a-1:\tau+b-a)  = \left[\begin{array}{c|c|c|c|c|c}
	& \multirow{5}{*}{$C$} & & &\\
	\cline{1-1} \cline{3-6} & & Z_2^* & & &\\
		 & & 0 & & \\
		 	\cline{1-1} \cline{3-6} & & & Z_2 & & \\
	\cline{1-1} \cline{3-6} & & & &\ddots \\  \cline{1-1} \cline{3-6}
	Z_1 &  &  & & & Z_2
\end{array}\right]
\eean\\

\emph{Random Erasure Recovery:} Suppose there exists an $A \subseteq [(\ell-1)a+a-1:\tau+b-a]$ such that $|A| \le a$ and $\sum_{j \in A} u_j\underline{h}_j=0$, $u_j \in \mathbb{F}_q$. The Cauchy columns are the only columns with non-zero support in rows $[0,a-2] \cup {2a-2}$. No collection of $\le a$ columns from Cauchy part can combine to form these $a$ zeros, hence $A \cap [b:\tau] = \phi$. Now due to MDS property of $Z$ and the support structure, it is not possible for $\le a$ columns from the remaining part to be linearly dependent. Hence, no such $A$ exists. \\  

\emph{Burst Erasure Recovery:}  Consider any submatrix $\hat{H}$ formed by $b$ consecutive columns of $H(0:b-1,~(\ell-1)a+a-1:\tau+b-a)$. It can be seen that the submatrix of $\hat{H}$ formed by Cauchy columns and rows where non-cauchy columns have no support is square and hence invertible. The condition $\tau+1-b \ge a$ ensures that more than $a$ columns containing entries from same $Z$ matrix are not part of $\hat{H}$. Now, it can be easily argued that $\hat{H}$ is invertible using support structure of $\hat{H}$ and properties of $Z$.   
\end{enumerate}
\eprf

\bibliographystyle{IEEEtran}
\bibliography{simple_streaming}	

\begin{thebibliography}{1}
\providecommand{\url}[1]{#1}
\csname url@samestyle\endcsname
\providecommand{\newblock}{\relax}
\providecommand{\bibinfo}[2]{#2}
\providecommand{\BIBentrySTDinterwordspacing}{\spaceskip=0pt\relax}
\providecommand{\BIBentryALTinterwordstretchfactor}{4}
\providecommand{\BIBentryALTinterwordspacing}{\spaceskip=\fontdimen2\font plus
\BIBentryALTinterwordstretchfactor\fontdimen3\font minus
  \fontdimen4\font\relax}
\providecommand{\BIBforeignlanguage}[2]{{%
\expandafter\ifx\csname l@#1\endcsname\relax
\typeout{** WARNING: IEEEtran.bst: No hyphenation pattern has been}%
\typeout{** loaded for the language `#1'. Using the pattern for}%
\typeout{** the default language instead.}%
\else
\language=\csname l@#1\endcsname
\fi
#2}}
\providecommand{\BIBdecl}{\relax}
\BIBdecl

\bibitem{MartSunTIT04}
E.~Martinian and C.~W. Sundberg, ``Burst erasure correction codes with low
  decoding delay,'' \emph{{IEEE} Trans. Inf. Theory}, vol.~50, no.~10, pp.
  2494--2502, 2004.

\bibitem{MartTrotISIT07}
E.~Martinian and M.~Trott, ``{Delay-Optimal Burst Erasure Code Construction},''
  in \emph{{Proc. Int. Symp. Inf. Theory, Nice, France, June 24-29,
  2007}}.\hskip 1em plus 0.5em minus 0.4em\relax {IEEE}, pp. 1006--1010.

\bibitem{BadrPatilKhistiTIT17}
A.~Badr, P.~Patil, A.~Khisti, W.~Tan, and J.~G. Apostolopoulos, ``{Layered
  Constructions for Low-Delay Streaming Codes},'' \emph{{IEEE} Trans. Inf.
  Theory}, vol.~63, no.~1, pp. 111--141, 2017.

\bibitem{NikDeepPVK}
M.~N. Krishnan, D.~Shukla, and P.~V. Kumar, ``Low field-size, rate-optimal
  streaming codes for channels with burst and random erasures,'' \emph{{IEEE}
  Trans. Inf. Theory, Early Access}, 2020.

\bibitem{FongKhistiTIT19}
S.~L. Fong, A.~Khisti, B.~Li, W.~Tan, X.~Zhu, and J.~G. Apostolopoulos,
  ``Optimal streaming codes for channels with burst and arbitrary erasures,''
  \emph{{IEEE} Trans. Inf. Theory}, vol.~65, no.~7, pp. 4274--4292, 2019.

\bibitem{NikPVK}
M.~N. Krishnan and P.~V. Kumar, ``{Rate-Optimal Streaming Codes for Channels
  with Burst and Isolated Erasures},'' in \emph{{Proc. Int. Symp. Inf. Theory,
  Vail, CO, USA, June 17-22, 2018}}.\hskip 1em plus 0.5em minus 0.4em\relax
  {IEEE}, pp. 1809--1813.

\bibitem{KhistiExplicitCode}
E.~Domanovitz, S.~L. Fong, and A.~Khisti, ``An explicit rate-optimal streaming
  code for channels with burst and arbitrary erasures,'' 2019,
  arXiv:1904.06212.

\bibitem{simple}
M.~N. Krishnan, V.~Ramkumar, M.~Vajha, and P.~V. Kumar, ``Simple streaming
  codes for reliable, low-latency communication,'' \emph{{IEEE} Communications
  Letters}, vol.~24, no.~2, pp. 249--253, 2020.

\bibitem{ecc_Mac_Slo}
F.~J. MacWilliams and N.~J.~A. Sloane, \emph{The theory of error-correcting
  codes}.\hskip 1em plus 0.5em minus 0.4em\relax Elsevier, 1977, vol.~16.

\end{thebibliography}

\end{document}